\documentclass[conference,a4paper]{IEEEtran}
\IEEEoverridecommandlockouts
\usepackage{cite}
\usepackage{amsmath,amssymb,amsfonts}
\usepackage{algorithmic}
\usepackage{graphicx}
\usepackage{textcomp}
\usepackage{xcolor}
\usepackage{placeins}
\def\BibTeX{{\rm B\kern-.05em{\sc i\kern-.025em b}\kern-.08em
    T\kern-.1667em\lower.7ex\hbox{E}\kern-.125emX}}
\begin{document}

\title{Lunar Surface Receiver Clock Synchronization Using GNSS and Satellite Nadir Angle
\thanks{This work was supported in part by Grant RS-2024-00407003 from the ``Development of Advanced Technology for Terrestrial Radionavigation System'' project, funded by the Ministry of Oceans and Fisheries, Republic of Korea;
in part by the National Research Foundation of Korea (NRF), funded by the Korean government (Ministry of Science and ICT), under Grant RS-2024-00358298; 
in part by the Future Space Navigation and Satellite Research Center through the NRF, funded by the Ministry of Science and ICT (MSIT), Republic of Korea, under Grant 2022M1A3C2074404; 
in part by the Basic Science Research Program through the NRF, funded by the Ministry of Education, Republic of Korea, under Grant RS-2025-25425504; 
and in part by the MSIT, Korea, under the Information Technology Research Center (ITRC) support program supervised by the Institute of Information \& Communications Technology Planning \& Evaluation (IITP) under Grant IITP-2024-RS-2024-00437494.
}
}

\author{\IEEEauthorblockN{Minju Han} 
\IEEEauthorblockA{\textit{School of Integrated Technology} \\
\textit{Yonsei University}\\
Incheon, Korea \\
hanmju@yonsei.ac.kr} 
\and
\IEEEauthorblockN{Seunghyeon Park} 
\IEEEauthorblockA{\textit{School of Integrated Technology} \\
\textit{Yonsei University}\\
Incheon, Korea \\
seunghyeon.park@yonsei.ac.kr} 
\and
\IEEEauthorblockN{Joon Hyo Rhee${}^{*}$} 
\IEEEauthorblockA{\textit{Korea Research Institute of Standards and Science} \\
Daejeon, Korea \\
jh.rhee@kriss.re.kr}
{\small${}^{*}$ Corresponding author}
}

\maketitle

\begin{abstract}
The use of signals from Earth-orbiting Global Navigation Satellite System (GNSS) constellations for positioning, navigation, and timing on the lunar surface was recently demonstrated by the Lunar GNSS Receiver Experiment (LuGRE) aboard Blue Ghost Mission 1.
For receiver clock synchronization using GNSS, Kalman filter performance depends on the satellite-specific measurement uncertainty used to construct the measurement noise covariance.
Terrestrial measurement weighting commonly relies on receiver elevation angle, but the elevation angles in the analyzed LuGRE data are confined to 24.2$^\circ$--31.5$^\circ$, providing limited discrimination of measurement quality.
In contrast, satellite nadir angles span 12.1$^\circ$--66.6$^\circ$, and the measurement variability increases substantially at small nadir angles.
Based on this observation, we propose a measurement weighting model that combines satellite nadir angle with $C/N_0$ for lunar surface receiver clock synchronization.
The proposed model is evaluated against three weighting methods adopted from the literature using seven LuGRE operation periods (OPs).
It achieves the lowest receiver clock bias root mean square error (RMSE) in six of the seven OPs and an overall RMSE of 5.78~m, which is 19.5\% lower than that of the best comparison model.
The proposed model also provides the lowest clock bias prediction RMSE under simulated measurement outages.
\end{abstract}


\section{Introduction}
\label{sec:introduction}

Global Navigation Satellite System (GNSS) has become a key infrastructure for positioning, navigation, and timing (PNT), with applications extending from terrestrial and low Earth orbit environments to high altitude and cislunar navigation \cite{Bauer06:GPS, Jing15:SSV, Winternitz17:Global}.
Recent PNT research has addressed positioning and timing reliability in dense urban environments \cite{Kim26:Zonotope, Kim25:Set}, alternative terrestrial radionavigation systems \cite{Park25:Toward, Kang25:Enhancing}, and lunar navigation systems \cite{Park26:LNSS}.
In parallel, the direct use of signals transmitted by Earth-orbiting GNSS satellites for lunar operations has been investigated in a number of studies \cite{Konitzer24:Science, Delepaut19:System, Delepaut20:Use, Winternitz19:GPS, Giordano22:Lunar, Peters26:Analysis}.
Such signals can provide an additional source of position and timing information for lunar missions and complement dedicated lunar navigation infrastructure \cite{Israel20:LunaNet, Bhamidipati23:Satellite}.
The Lunar GNSS Receiver Experiment (LuGRE) aboard Blue Ghost Mission 1 (BGM1) \cite{Parker22:Lunar} recently demonstrated this capability by receiving and processing Global Positioning System (GPS) and Galileo signals on the lunar surface \cite{Parker26:GNSS}.

Accurate receiver clock synchronization is an important component of PNT using GNSS, particularly when measurements are weak, intermittent, or geometrically constrained.
Previous studies have also investigated clock synchronization and clock state prediction under limited GNSS availability \cite{Park24:CSAC, Park26:LNSS}.
In a Kalman filter-based clock estimator, the measurement update depends on the uncertainty assigned to each satellite observation, which determines the measurement noise covariance used by the filter.
Terrestrial GNSS stochastic models commonly characterize this uncertainty using receiver elevation angle, carrier-to-noise density ratio ($C/N_0$), or combinations of these quantities \cite{Groves13:Height, Li22:Combined, Li16:Elevation}.
Elevation-based weighting is effective in many terrestrial environments because measurements received at low elevation angles are generally more susceptible to atmospheric effects and multipath \cite{Kim25:Enhancing}.

The receiver elevation geometry is substantially different on the lunar surface.
Because the Earth--Moon distance is much larger than the spatial extent of the GNSS constellations, Earth-orbiting GNSS satellites occupy a narrow angular region when viewed from a lunar receiver \cite{Delepaut19:System, Delepaut20:Use, Bhamidipati23:Satellite}.
Here, the satellite nadir angle denotes the angle, measured at the transmitting GNSS satellite, between the satellite-to-Earth-center direction and the satellite-to-receiver direction.
In the seven LuGRE operation periods (OPs) considered in this study, the receiver elevation angles were confined to 24.2$^\circ$--31.5$^\circ$, whereas the corresponding satellite nadir angles spanned 12.1$^\circ$--66.6$^\circ$.
More importantly, the LuGRE measurement variability showed no clear systematic dependence over the narrow elevation angle range, whereas it increased substantially at small satellite nadir angles.
These observations indicate that satellite nadir angle provides additional information for discriminating measurement quality under lunar surface GNSS reception conditions.

Based on these observations, we propose a measurement weighting model that combines satellite nadir angle with $C/N_0$ for lunar surface receiver clock synchronization.
The proposed model is evaluated using seven LuGRE OPs against three weighting methods adopted from the literature: equal weighting, $C/N_0$-based weighting, and elevation-based weighting.
The parameters of the comparison models are adopted directly from the corresponding literature, while the parameters of the proposed model are selected through leave-one-OP-out cross-validation.
The proposed model achieves the lowest receiver clock bias root mean square error (RMSE) in six of the seven OPs and reduces the overall RMSE by 19.5\% relative to the best comparison model.
As a complementary evaluation, clock bias prediction under simulated measurement outages is examined to assess the quality of the estimated clock states after measurement updates are removed.

\begin{figure}[t]
    \centering
    \includegraphics[width=0.7\linewidth]{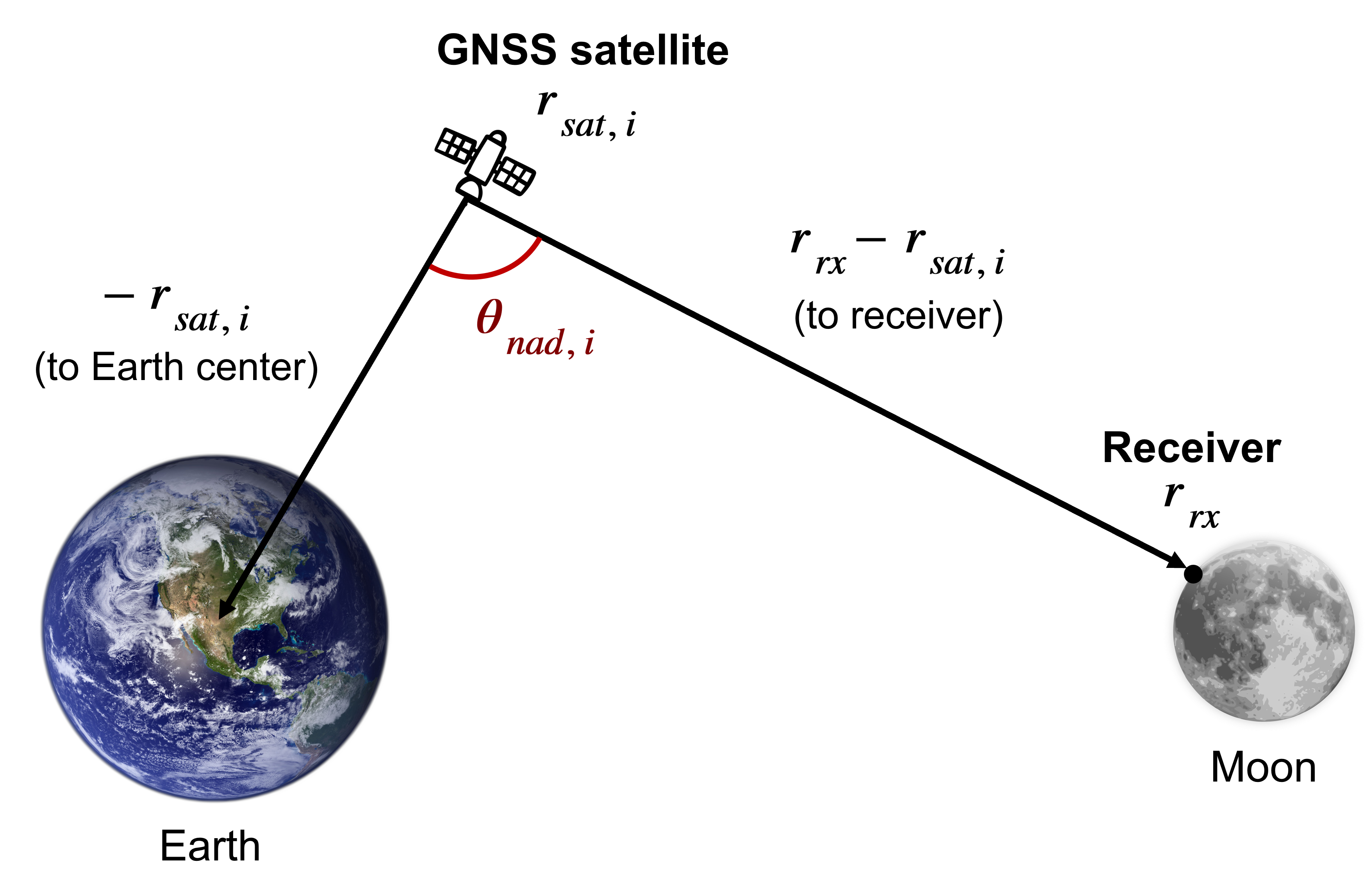}
    \caption{Geometric definition of the satellite nadir angle $\theta_{nad,i}$, measured between the satellite-to-Earth-center vector and the satellite-to-receiver vector (not to scale).}
    \label{fig:nadir_geometry}
\end{figure}

\section{Methodology}
\label{sec:methodology}

\subsection{Kalman Filter for Clock Synchronization}
\label{subsec:kf}

For satellite $i$ at epoch $k$, the receiver clock bias measurement is constructed from the code phase observation as \cite{Kaplan17:Understanding}
\begin{equation}
\begin{aligned}
z_{i,k} ={}& pr_{raw,i,k} - d_{gcpm,i,k} - \rho_{i,k} \\
&+ clk_{sat,i,k} + rel_{corr,i,k} - bias_{code,i,k},
\end{aligned}
\label{eq:measurement}
\end{equation}
where $pr_{raw,i,k}$ is the raw code phase observation, $d_{gcpm,i,k}$ denotes the plasma propagation delay correction based on the Global Core Plasma Model (GCPM) \cite{Gallagher00:Global}, $\rho_{i,k}$ is the geometric range between the satellite and receiver, $clk_{sat,i,k}$ is the satellite clock correction, $rel_{corr,i,k}$ is the periodic relativistic correction associated with the satellite orbital eccentricity \cite{Ashby03:Relativity}, and $bias_{code,i,k}$ is the code bias.
The relativistic correction, expressed in range units, is computed as
\begin{equation}
rel_{corr,i,k}
=
-\frac{2}{c}
\mathbf{r}_{sat,i,k}^{T}
\mathbf{v}_{sat,i,k},
\label{eq:relativistic}
\end{equation}
where $c$ is the speed of light, and $\mathbf{r}_{sat,i,k}$ and $\mathbf{v}_{sat,i,k}$ are the satellite position and velocity evaluated at the signal transmission time.
Because the periodic relativistic term is treated separately from the Wuhan University (WHU) precise clock product in the processing adopted in this study, it is explicitly included in (\ref{eq:measurement}).
The WHU multi-GNSS products provide GPS and Galileo satellite clocks in a common solution aligned with GPS Time (GPST) \cite{Guo16:Precise}, and the corresponding observable-specific code bias corrections \cite{Geng22:OSB} are applied to both constellations.
A separate GPS--Galileo inter-system bias is not estimated in this study.
An epoch-wise diagnostic shows a consistent constellation-dependent offset of approximately 2.5~m, which may include both receiver-dependent and residual propagation effects and is treated as part of the measurement error.
The resulting measurement can therefore be expressed as $z_{i,k}=b_k+\epsilon_{i,k}$, where $b_k$ denotes the receiver clock bias in meters relative to GPST and $\epsilon_{i,k}$ represents the measurement error.

At each epoch, the available satellite measurements are combined using inverse variance weights $\omega_{i,k}=\sigma_{i,k}^{-2}$ as
\begin{equation}
z_k =
\frac{\sum_i \omega_{i,k} z_{i,k}}
{\sum_i \omega_{i,k}}.
\label{eq:combine}
\end{equation}
Assuming independent measurement errors, the variance of the combined measurement is
$R_k=(\sum_i \omega_{i,k})^{-1}$,
where $\sigma_{i,k}$ denotes the measurement standard deviation assigned to satellite $i$.

The receiver clock state is estimated using a two-state Kalman filter \cite{Zucca05:Clock} with the state vector $\mathbf{x}_k=[b_k\;\;\dot{b}_k]^T$.
The state transition model is
\begin{equation}
\mathbf{x}_{k+1} = \mathbf{F}_k\mathbf{x}_k + \mathbf{w}_k, \qquad
\mathbf{F}_k =
\begin{bmatrix}
1 & \Delta t_k \\
0 & 1
\end{bmatrix},
\label{eq:state_model}
\end{equation}
and the measurement model is
\begin{equation}
z_k = \mathbf{H}\mathbf{x}_k + v_k, \qquad
\mathbf{H} = \begin{bmatrix}1 & 0\end{bmatrix},
\label{eq:measurement_model}
\end{equation}
where $\Delta t_k$ is the time interval between consecutive epochs, $\mathbf{w}_k$ is the process noise, and $v_k$ is the combined measurement noise with variance $R_k$.

The process noise covariance is determined from the clock noise parameters $q_{1c}^2$ and $q_{2c}^2$ \cite{Zucca05:Clock} as
\begin{equation}
\mathbf{Q}_k =
\begin{bmatrix}
q_{1c}^2\Delta t_k + \dfrac{q_{2c}^2\Delta t_k^3}{3}
&
\dfrac{q_{2c}^2\Delta t_k^2}{2}
\\[4pt]
\dfrac{q_{2c}^2\Delta t_k^2}{2}
&
q_{2c}^2\Delta t_k
\end{bmatrix}.
\label{eq:Q}
\end{equation}

Because the same state and process models are used for all compared methods, their primary difference lies in the satellite-specific measurement uncertainty $\sigma_{i,k}$ used to construct $R_k$.
The following subsection develops a measurement noise model suited to lunar GNSS observation geometry.

\subsection{Measurement Noise Modeling Using Nadir Angle}
\label{subsec:nadir_model}

Terrestrial GNSS measurement noise models commonly use receiver elevation angle as an indicator of measurement quality.
At lunar distances, however, the angular separation among Earth-orbiting GNSS satellites is highly compressed when viewed from the receiver.
Consequently, elevation angle provides limited geometric discrimination among simultaneously observed measurements.

The satellite nadir angle is defined as the angle between the satellite-to-Earth-center vector and the satellite-to-receiver vector, as illustrated in Fig.~\ref{fig:nadir_geometry}.
For notational simplicity, the epoch index $k$ is omitted from the nadir angle and measurement noise model expressions below.

The nadir angle is computed as
\begin{equation}
\theta_{nad,i}
=
\cos^{-1}
\left(
\frac{
(-\mathbf{r}_{sat,i})
\cdot
(\mathbf{r}_{rx}-\mathbf{r}_{sat,i})
}{
\lVert\mathbf{r}_{sat,i}\rVert
\lVert\mathbf{r}_{rx}-\mathbf{r}_{sat,i}\rVert
}
\right),
\label{eq:nadir}
\end{equation}
where $\mathbf{r}_{sat,i}$ and $\mathbf{r}_{rx}$ denote the Earth-centered satellite and receiver position vectors expressed in the same reference frame.

For space users above the GNSS constellations, received signal characteristics can vary with satellite-centered transmission geometry \cite{Marquis15:Antenna, Shehaj17:SSV, Menzione24:Galileo}.
The observed measurement variability can additionally reflect propagation effects along the corresponding signal path.

To characterize satellite-specific measurement variability without using the reference clock solution or Kalman filter estimate, the deviation of each measurement from the unweighted mean of the simultaneously available measurements is defined as
\begin{equation}
e_{i,k}
=
\left(
z_{i,k}
-
\frac{1}{N_k}\sum_{j=1}^{N_k}z_{j,k}
\right)
\sqrt{\frac{N_k}{N_k-1}},
\label{eq:residual}
\end{equation}
where $N_k$ is the number of available satellite measurements at epoch $k$.
The correction factor compensates for the variance reduction caused by including each measurement in the epoch-wise mean.
Measurement variability within each angular bin is then quantified using 1.4826 times the median absolute deviation (MAD), which provides a robust estimate of standard deviation.

\begin{figure}[t]
    \centering
    \includegraphics[width=1.0\linewidth]{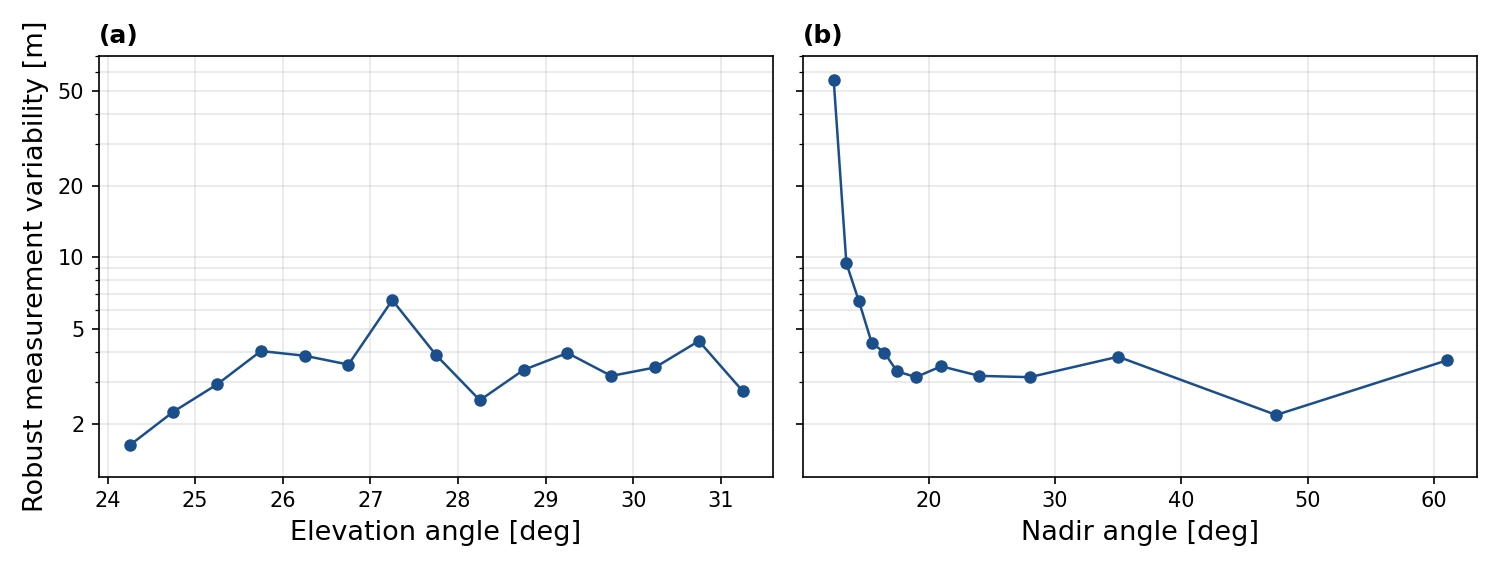}
    \caption{Robust measurement variability versus (a) receiver elevation angle and (b) satellite nadir angle across the seven LuGRE OPs (logarithmic vertical scale).}
    \label{fig:residual_angle}
\end{figure}

Applying this characterization to the seven LuGRE OPs described in Section~\ref{subsec:dataset} yields the relationships shown in Fig.~\ref{fig:residual_angle}.
The analysis uses 295,604 GPS L5 and Galileo E5a measurements from epochs with $N_k\geq2$; bins containing fewer than 300 measurements are excluded.
Elevation angles are grouped into 0.5$^\circ$ bins, whereas variable-width nadir angle bins are used to account for the nonuniform sample density.

As shown in Fig.~\ref{fig:residual_angle}(a), the measurement variability shows no clear systematic dependence over the narrow elevation angle range.
In contrast, Fig.~\ref{fig:residual_angle}(b) shows a pronounced increase in measurement variability at small nadir angles.
This behavior motivates the use of satellite nadir angle as an additional indicator of measurement uncertainty.

Based on the observed measurement variability, the satellite-specific measurement standard deviation is modeled as
\begin{equation}
\sigma_i
=
\sigma_0
+
A\exp\left(-\frac{\theta_{nad,i}-\theta_{ref}}{s_\theta}\right)
10^{-((C/N_0)_i-C_{ref})/20},
\label{eq:proposed_general}
\end{equation}
where $\sigma_0$ is the measurement noise floor, $A$ determines the magnitude of the nadir-dependent term, $\theta_{ref}$ and $s_\theta$ control its angular dependence, and $C_{ref}$ is the reference $C/N_0$ level in dB-Hz.
The nadir-dependent term assigns larger uncertainty to measurements at small nadir angles, while the $C/N_0$ term accounts for measurement quality associated with signal strength.
The numerical parameters of (\ref{eq:proposed_general}) are determined using the cross-validation procedure described in Section~\ref{subsec:models}.

\section{Experiment}
\label{sec:experiment}

\subsection{LuGRE Dataset and Reference Solution}
\label{subsec:dataset}

This study uses GPS L5 and Galileo E5a observations collected by the LuGRE payload on the lunar surface after the BGM1 landing \cite{Parker26:GNSS}.
Seven OPs are analyzed: OP38\_0, OP40\_0, OP74\_0, OP76\_0, OP77\_0, OP77\_1, and OP78\_1.
Their durations range from tens of minutes to more than 20 hours, providing a total of 400,821 satellite observations.

Satellite positions were obtained from WHU precise orbit products \cite{Guo16:Precise}, while satellite clock corrections and code biases were obtained from the associated WHU clock and observable-specific bias products \cite{Geng22:OSB}.
For each observation, the satellite position was linearly interpolated to the signal transmission time, approximated as $t_{tx}=t_{rx}-pr_{raw}/c$.
The corresponding satellite velocity used in (\ref{eq:relativistic}) was obtained by finite differencing successive precise orbit positions at 1~s intervals.

The lunar receiver position was generated from the known BGM1 landing site coordinates of $18.5623^\circ$N and $61.8103^\circ$E \cite{Banks26:BGM1}, expressed in the Moon-fixed IAU\_MOON frame.
The site radius was set to 1733.75~km, corresponding to a site elevation of 3.650~km below the mean lunar radius of 1737.4~km.
At each epoch, the landing site position was transformed to J2000 using the lunar orientation model \cite{Archinal18:IAU}.
The Moon's geocentric position was obtained from the DE440s ephemeris \cite{Park21:DE440}, after which the resulting Earth-centered J2000 position was transformed to the International Terrestrial Reference Frame 1993 (ITRF93) Earth-fixed frame using NASA's SPICE toolkit \cite{Acton96:Ancillary, Acton18:Look}.
The resulting epoch-dependent Earth-centered receiver position was used to compute the geometric range in (\ref{eq:measurement}) and the nadir angle in (\ref{eq:nadir}).
For elevation angle computation, the receiver local vertical was defined with respect to the center of the Moon.

Across the seven OPs, the elevation angles were confined to 24.2$^\circ$--31.5$^\circ$, with a mean of 27.9$^\circ$ and a standard deviation of 1.4$^\circ$.
In contrast, the nadir angles spanned 12.1$^\circ$--66.6$^\circ$, with a mean of 21.2$^\circ$ and a standard deviation of 7.4$^\circ$.
The substantially wider nadir angle range provides greater geometric variation for satellite-specific measurement noise modeling.

To reduce direct information overlap between the estimated and reference clock solutions, the Kalman filter measurements in (\ref{eq:measurement}) are constructed using code phase observations only.
The reference clock solution used for performance evaluation is reconstructed using both code and carrier phase observations.
The carrier phase observations are segmented into continuous arcs, leveled to the code observations, and jointly adjusted, with the absolute clock level anchored by the code measurements.
This separation reduces the risk that a measurement weighting scheme is favored simply because it resembles the combination procedure used to generate the reference solution.

\subsection{Comparison Models and Proposed Configuration}
\label{subsec:models}

Four measurement noise models are evaluated using the same clock state estimation framework.
The first three models are adopted directly from the literature without retuning their parameters to the LuGRE data, whereas the parameters of the proposed model are determined through leave-one-OP-out cross-validation.

\noindent\textbf{(i) Equal weighting:}
The same standard deviation of 5.67~m is assigned to all available satellite measurements \cite{Hutsell95:Kalman}.
Accordingly, for $N_k$ independent measurements available at epoch $k$,
\begin{equation}
R_k = \frac{5.67^2}{N_k}.
\label{eq:equal}
\end{equation}

\noindent\textbf{(ii) $C/N_0$-based weighting:}
The measurement variance is determined as a function of $C/N_0$ using the model in \cite{Hartinger99:Variances}, as adopted in \cite{Groves13:Height}:
\begin{equation}
\sigma_i^2 = 1.1\times10^4\,10^{-(C/N_0)_i/10},
\label{eq:cn0}
\end{equation}
and the measurements are combined according to (\ref{eq:combine}).

\noindent\textbf{(iii) Elevation-based weighting:}
The measurement standard deviation follows the elevation-dependent model adopted in \cite{Groves13:Height}:
\begin{equation}
\sigma_i = 0.13 + 0.56\exp\left(-\frac{\theta_{el,i}}{10^\circ}\right),
\label{eq:elevation}
\end{equation}
where $\theta_{el,i}$ denotes the elevation angle of satellite $i$.

\noindent\textbf{(iv) Proposed nadir- and $C/N_0$-based weighting:}
The proposed model follows (\ref{eq:proposed_general}), whose functional form is motivated by the measurement variability characterized in Section~\ref{subsec:nadir_model}.
Its numerical parameters are determined through a grid search with leave-one-OP-out cross-validation.
The grid comprises $\sigma_0 \in$ \{0.5, 1, 2\}~m, $A \in$ \{20, 47, 100\}~m, and $s_\theta \in$ \{0.5$^\circ$, 0.8$^\circ$, 1.2$^\circ$\}, with the sum of the per-OP RMSE values over the six training OPs used as the selection criterion.
The reference values $\theta_{ref}$ = 11$^\circ$ and $C_{ref}$ = 30~dB-Hz are fixed and are not tuned.
For each fold, one OP is excluded, the parameters are selected using the remaining six OPs, and the selected model is evaluated on the held-out OP.
All seven folds selected the same parameter combination, which was also identical to that selected using the full dataset, yielding
\begin{equation}
\sigma_i
=
0.5
+
20\exp\left(-\frac{\theta_{nad,i}-11^\circ}{1.2^\circ}\right)
10^{-((C/N_0)_i-30)/20}.
\label{eq:proposed}
\end{equation}

Consequently, the held-out results obtained through leave-one-OP-out cross-validation are numerically identical to those obtained using the parameter set selected using the full dataset.

\subsection{Evaluation Procedure}
\label{subsec:evaluation}

All four measurement noise models are applied to the same two-state clock Kalman filter described in Section~\ref{subsec:kf}.
The LuGRE receiver employs a voltage-controlled temperature-compensated crystal oscillator \cite{Konitzer22:Navigation}.
The process noise parameters are computed from representative temperature-compensated crystal oscillator noise coefficients reported in \cite{Hsu15:Multipath},
\begin{equation}
q_{1c}^2=0.0090~\mathrm{m^2/s},
\qquad
q_{2c}^2=0.0355~\mathrm{m^2/s^3}.
\end{equation}

The same filter settings are used for all models, and measurement gating and filter reinitialization are disabled to maintain identical comparison conditions. 
The filter is initialized at the first epoch of each OP with $b_0=z_0$, $\dot{b}_0=0$, and a diffuse initial covariance of $10^8$~m$^2$ for bias and $10^4$~(m/s)$^2$ for drift, reflecting the absence of prior information.
Clock synchronization performance is evaluated in terms of the receiver clock bias RMSE for each OP and for the pooled dataset.
For the proposed model, the per-OP RMSE values therefore also correspond to the held-out results of the leave-one-OP-out cross-validation.

As a complementary evaluation, clock bias prediction under simulated measurement outages is performed to examine the quality of the estimated clock states after measurement updates are removed.
Within each OP, outage start epochs are selected at intervals of 30~min, excluding the 60~s immediately following a signal gap.
A total of 142 outage start epochs are evaluated across the seven OPs, with overlapping prediction windows treated independently.
From each start epoch, the estimated bias and drift states are propagated for up to one hour without measurement updates and compared with the reference clock solution.
Prediction errors are evaluated over five elapsed time intervals: 0--60~s, 60--300~s, 300--900~s, 900--1800~s, and 1800--3600~s.

\section{Results}
\label{sec:results}

\subsection{Clock Bias Estimation Results}
\label{subsec:results}

Table~\ref{tab:overall} compares the receiver clock bias RMSE of the four measurement noise models across the seven OPs and the pooled dataset.

\begin{table}[t]
\centering
\renewcommand{\arraystretch}{1.2}
\caption{Receiver Clock Bias RMSE for Four Measurement Weighting Methods [m]}
\label{tab:overall}
\begin{tabular}{lcccc}
\hline
& \multicolumn{4}{c}{Measurement Weighting Method} \\
\cline{2-5}
OP & Equal & $C/N_0$ & Elevation & Proposed \\
\hline
OP38\_0 & 7.38 & 6.21 & 7.02 & \textbf{5.19} \\
OP40\_0 & 9.67 & \textbf{5.10} & 7.64 & 5.16 \\
OP74\_0 & 7.01 & 8.00 & 6.51 & \textbf{5.13} \\
OP76\_0 & 11.69 & 8.07 & 11.05 & \textbf{7.22} \\
OP77\_0 & 6.56 & 6.26 & 5.67 & \textbf{5.63} \\
OP77\_1 & 2.90 & 2.90 & 2.88 & \textbf{2.87} \\
OP78\_1 & 1.16 & 1.13 & 1.12 & \textbf{1.06} \\
\hline
\textbf{Overall} & 8.73 & 7.18 & 8.13 & \textbf{5.78} \\
\hline
\end{tabular}
\end{table}
The proposed model achieved the lowest receiver clock bias RMSE in six of the seven OPs as well as for the pooled dataset.
The only exception was OP40\_0, where the $C/N_0$-based model achieved 5.10~m compared with 5.16~m for the proposed model.
The overall RMSE was 5.78~m, compared with 8.73~m for equal weighting, 7.18~m for $C/N_0$-based weighting, and 8.13~m for elevation-based weighting.
Compared with the best comparison model ($C/N_0$-based weighting), the proposed model reduced the overall RMSE by 19.5\%.

The per-OP RMSE variation is also consistent with differences in satellite availability and observation geometry.
For example, OP76\_0, which yields the largest RMSE under the proposed model, has substantially higher fractions of single-satellite epochs and small-nadir-angle observations than OP78\_1, which yields the lowest RMSE.
These results indicate that the proposed model provides more effective satellite-specific measurement weighting than the evaluated terrestrial weighting models under the LuGRE lunar surface geometry.

\subsection{Clock Bias Prediction Under Measurement Outages}
\label{subsec:outage_results}

Table~\ref{tab:outage} summarizes the clock bias prediction RMSE as a function of elapsed time after the beginning of each simulated measurement outage.

The proposed model achieved the lowest prediction RMSE in all five elapsed time intervals.
The improvement was most pronounced immediately after the beginning of an outage.
During the first 60~s, the proposed model achieved an RMSE of 12.78~m, which was 29.0\% lower than the 18.00~m obtained with the best comparison model, the $C/N_0$-based method.

Because the measurement noise covariance affects the Kalman filter only during measurement updates, its influence on the outage prediction is indirect.
The different measurement noise models produce different bias and drift state estimates at the beginning of each outage, after which the states are propagated without further measurement correction.
As the prediction horizon increases, the relative differences among the four models decrease substantially, indicating that long-term prediction error becomes increasingly dominated by clock state propagation rather than by the preceding measurement weighting.
This behavior is consistent with the limitation of propagating the clock state under a constant drift model over long measurement outages \cite{Winternitz19:GPS}.

\begin{table}[t]
\centering
\renewcommand{\arraystretch}{1.2}
\caption{Clock Bias Prediction RMSE Under Simulated Measurement Outages [m]}
\label{tab:outage}
\begin{tabular}{lcccc}
\hline
& \multicolumn{4}{c}{Measurement Weighting Method} \\
\cline{2-5}
Elapsed Time & Equal & $C/N_0$ & Elevation & Proposed \\
\hline
0--60~s & 25.49 & 18.00 & 31.89 & \textbf{12.78} \\
60--300~s & 217.56 & 202.27 & 232.58 & \textbf{190.57} \\
300--900~s & 1676.62 & 1649.17 & 1674.99 & \textbf{1614.18} \\
900--1800~s & 6681.29 & 6643.70 & 6686.37 & \textbf{6582.90} \\
1800--3600~s & 18839.22 & 18798.74 & 18908.46 & \textbf{18738.83} \\
\hline
\end{tabular}
\end{table}

\section{Conclusion}
\label{sec:conclusion}

This study developed a measurement weighting model that incorporates satellite nadir angle and $C/N_0$ for lunar surface receiver clock synchronization using GNSS.
Analysis of seven LuGRE OPs showed that receiver elevation angle was confined to a narrow range and exhibited no clear systematic relationship with measurement variability.
In contrast, measurement variability increased substantially at small satellite nadir angles.
Based on this observation, the proposed model assigns satellite-specific measurement uncertainty using satellite nadir angle and $C/N_0$.

The proposed model achieved the lowest receiver clock bias RMSE in six of the seven OPs.
Its overall RMSE was 5.78~m, representing a 19.5\% reduction relative to the best comparison model.
The proposed model also achieved the lowest clock bias prediction RMSE in all evaluated intervals under simulated measurement outages, with the largest relative improvement observed during the first 60~s.

These results indicate that satellite observation geometry can provide useful information for GNSS measurement weighting when receiver elevation angles are confined to a narrow range at lunar distances.
Future work should further investigate the physical mechanisms underlying the increased measurement variability at small nadir angles.
The proposed model should also be evaluated using additional lunar GNSS datasets and receiver configurations to assess the transferability of the selected parameters.

\section*{Acknowledgment}
\label{sec:acknowledgment}

Generative AI (ChatGPT, OpenAI) was used solely to assist with grammar and language improvements during the manuscript preparation process.  
No content, ideas, data, or citations were generated by AI.  
All technical content, methodology, analysis, and conclusions were written and verified solely by the authors.

\bibliographystyle{IEEEtran}
\bibliography{mybibfile, IUS_publications}

\vspace{12pt}

\end{document}